\documentclass[twocolumn,superscriptaddress,preprintnumbers,amsmath,amssymb,aps,pre]{revtex4}
\usepackage{graphicx}% Include figure files
\usepackage{dcolumn}% Align table columns on decimal point
\usepackage{bm}% bold math
\begin{document}
\title{On the anomalous {changes of seismicity and} geomagnetic field prior to the 2011 $M_w$ 9.0 Tohoku earthquake}

\author{E. S. Skordas}
\affiliation{Solid State Section and Solid Earth Physics
Institute, Physics Department, University of Athens,
Panepistimiopolis, Zografos 157 84, Athens, Greece}
\author{N. V. Sarlis}
\affiliation{Solid State Section and Solid Earth Physics
Institute, Physics Department, University of Athens,
Panepistimiopolis, Zografos 157 84, Athens, Greece}

\begin{abstract}
Xu et al. [J. Asian Earth Sci. {\bf 77}, 59-65
(2013)] It has just been reported that approximately 2 months
prior to the $M_w$9.0 Tohoku earthquake that occurred in Japan on
11 March 2011 anomalous variations of the geomagnetic field have
been observed in the vertical component at a measuring station
about 135 km from the epicenter for about 10 days (4 to 14 January
2011). Here, we show that this observation is in striking
agreement with independent recent results obtained from natural
time analysis of seismicity in Japan. In particular, this analysis
has revealed that an unprecedented minimum of the order parameter
fluctuations of seismicity was observed around 5 January 2011,
thus pointing to the initiation at that date of a strong
precursory Seismic Electric Signals activity accompanied by the
anomalous geomagnetic field variations. Starting from this date,
natural time analysis of the subsequent seismicity indicates that
a strong mainshock was expected in a few days to one week after
08:40 LT on 10 March 2011.
\end{abstract}

%Uncomment for PACS numbers title message
%\pacs{91.30.-f,05.40.-a}
% Keywords required only for MST, PB, PMB, PM, JOA, JOB? %
\maketitle
\section{Introduction}\label{sec1}
Several papers have reported various electromagnetic signals being
detected before large earthquakes {in Greece
\citep{VAR84A,VAR84B,JAP2,EPL12}, Japan \citep{UYE02,UYE09,UYE09B},
California} \citep[e.g.,][see also \citealt{BLE10} and references
therein]{FRA90,BER91} {and
China\citep{nhessHuang2011,Huang11}}. \citet{Xu2013} just reported
unusual behavior of geomagnetic diurnal variations prior to the
Tohoku earthquake of magnitude $M_w$9.0 that occurred in Japan on
11 March 2011. They computed ratios of diurnal variations range
between the target station Esashi (ESA) located at about 135 km
from the epicenter and the remote reference station Kakioka (KAK)
about 302 km distant to the epicenter. Their results showed a
clear anomaly exceeding the statistical threshold in the vertical
component about 2 months before the earthquake occurrence. The
original records of geomagnetic fields of the ESA station also
exhibited anomalous behavior for about 10 days (4 to 14 January
2011) in the vertical component approximately 2 months before the
$M_w$9 earthquake. It is the objective of this short paper to show
that these findings of \citet{Xu2013} are in full accord with
recent results published independently \citep{TECTO13,PNAS13} on
the basis of the analysis of seismicity of Japan in a new time
domain -termed natural time \citep{NAT02}- which reveals some
dynamic features hidden in the time series of complex systems
\citep{ABE05,SPRINGER}.

Natural time analysis \citep{NAT01,NAT02,NAT03B,NAT03A} has found
applications in diverse fields and the relevant results have been
compiled by \citet{SPRINGER}. In the case of seismicity, in a time
series comprising $N$ earthquakes, the natural time $\chi_k = k/N$
serves as an index for the occurrence of the $k$-th earthquake. In
natural time analysis we study this index in conjunction with the
energy $Q_k$ released during the $k$-th earthquake of magnitude
$M_k$, i.e., the pair $(\chi_k, Q_k)$. Alternatively, one employs
the pair $(\chi_k,p_k)$, where

\begin{equation}
p_k=\frac{Q_k}{\sum_{n=1}^NQ_n}
\end{equation}
denotes the normalized energy released during the $k$-th
earthquake. It has been found
\citep{NAT01,NAT03B,NAT03A,NAT05C,SPRINGER} that the variance of
$\chi$ weighted for $p_k$, designated by $\kappa_1$ given by

\begin{equation}\label{kappa1}
\kappa_1=\sum_{k=1}^N p_k (\chi_k)^2- \left(\sum_{k=1}^N p_k
\chi_k \right)^2,
\end{equation}
plays a prominent role in identifying when a complex system
approaches the critical point.

Since the observed earthquake scaling laws \citep[e.g.
see][]{TUR97} are widely accepted to indicate the existence of
phenomena closely associated with the proximity of the system to a
critical point \citep[e.g.,][]{HOL06,SOR00,CAR94,XIA08},
\citet{NAT05C} took the view that earthquakes are
(non-equilibrium) critical phenomena and argued that the quantity
$\kappa_1$ given by Eq. (\ref{kappa1}) can be considered as an
order parameter for seismicity. It has been found
\citep{NAT01,SAR08,PNAS,SPRINGER} that a mainshock occurs in a few
days to one week after the $\kappa_1$ value is recognized to have
approached 0.07 in the natural time analysis of the seismicity
subsequent to the initiation of a Seismic Electric Signals (SES)
activity. This is made on the premise that the initiation of an
SES activity marks the time when the system enters the critical
regime based on the following grounds: The SES, which are low
frequency transient changes of the electric field of the Earth
that precede earthquakes and their physical properties enable the
determination of the epicenter and the magnitude of an impending
earthquake \citep[]{VAR84A,VAR84B,VAR91}, are probably generated
by means of the so called pressure stimulated polarization
currents (PSPC) model proposed by \citet{VARBOOK} \citep[see
also][]{VAR93}. This model, which motivated the SES research,
makes use of the widely accepted concept that the stress gradually
increases in the future focal region of an EQ. When this stress
reaches a {\em critical} value, a {\em cooperative} orientation of
the electric dipoles (which are anyhow present in the focal area
due to lattice imperfections in the ionic constituents of the
rocks, e.g., see \cite{VARALEX80,VARALEX81}) is attained. This leads to the emission of a transient
electric signal that constitutes an SES (the {\em cooperativity}
is a hallmark of criticality). The validity of this SES generation
mechanism is strengthened by the finding that the up to date
experimental data of SES activities have been shown to exhibit
infinitely ranged temporal correlations
\citep{NAT03A,NAT03B,NAT09V}, which conforms with the aspect of
{\em critical} dynamics. As pointed out by \citet{UYE09B}, the
PSPC model is unique among other models in that SES would be
generated spontaneously during the gradual increase of stress
without requiring any sudden change of stress such as
microfracturing{\citep{MOL95} \citep[or
faulting][]{Huang02,Ren12}}.

The up to date observations of the magnetic field variations
accompanying the SES activities have shown that are clearly
detectable at distances of the order of $\sim 100$ km for strong
EQs, i.e., of magnitude 6.5 or larger \citep[e.g., see][]{VAR03}.
In addition, as demonstrated by detailed computations
\citep{SAR02}, these magnetic field changes mainly appear in the
$Z$ component as observed by \citet{Xu2013}. This will be further
discussed below.

\section{Natural time analysis of seismicity in Japan. Recent results and their relation with the findings of Xu et al. [11]}\label{sec2}
By analyzing the Japanese seismic catalog in natural time, and
employing a sliding natural time window comprising the number of
events that would occur in a few months, the following results
have been recently published:

\citet{TECTO13} found that the fluctuations of the order parameter
$\kappa_1$ of seismicity exhibit a clearly detectable minimum
approximately at the time of the initiation of the pronounced SES
activity observed by \citet{UYE02,UYE09} almost two months before
the onset of the volcanic-seismic swarm activity in 2000 in the
Izu Island region, Japan. This reflects that presumably the same
physical cause led to both effects observed, i.e, the emission of
the SES activity and the change of the correlation properties
between the earthquakes. This might be the case  when the stress
reaches a {\em critical} value, if we recall the PSPC model
(mentioned above in Section \ref{sec1}) for the SES generation. In
addition, \citet{TECTO13} reported that the aforementioned two
phenomena were found to be also linked in space.

Note that for the vast majority of major earthquakes in Japan the
almost simultaneous appearance of these minima with the initiation
of SES activities cannot be checked due to the lack of
geoelectrical data. In view of this lack of data, \citet{PNAS13}
proceeded to the analysis of the Japan seismic catalog in natural
time from 1 January 1984 to 11 March 2011, the day of the $M_w$9
Tohoku earthquake.They found that the fluctuations of the order
parameter of seismicity exhibited distinct minima a few months
before all the shallow earthquakes of magnitude 7.6 or larger that
occurred during this 27 year period in Japanese area. Among the
minima, the minimum before the $M_w$9 Tohoku earthquake observed
on $\sim$ 5 January 2011 was the deepest. \citet{PNAS13} ended
their conclusion by pointing out that ``the approximate
coincidence of the lead time of minima of the order parameter
fluctuations of seismicity with that of the SES activities may
help in understanding the physics of both phenomena''. This,
reflects that for the case of the aforementioned deepest minimum
of seismicity before the Tohoku earthquake observed on $\sim$5
January 2011, a strong SES activity should have been initiated on
the same date. Consequently, as mentioned in Section \ref{sec1},
anomalous magnetic field changes accompanying the electric field
variations of this SES activity should also initiate on this date,
i.e., 5 January 2011. It is this expectation which is strikingly
verified by \citet{Xu2013} who reported that anomalous magnetic
field variations at Esashi station initiated on $\sim$4 January
and lasted almost 10 days.

\section{Determination of the occurrence time of Tohoku $M_w$9.0
earthquake.}\label{sec3} The SES activities are considered as
mentioned above to occur when the focal zone enters the critical
stage. The date of the minimum ($\sim$5 January) of the order
parameter fluctuations of seismicity determined by \citet{PNAS13}
-which almost coincides with the starting date of the magnetic
field variations identified by \cite{Xu2013}- could be considered
as the time of  the initiation of the SES activity. Assuming that
this is the case, we started the computation of the $\kappa_1$
values of seismicity from this time in each of all the subareas
\citep{SPRINGER,SAR08} in a region $3^0 \times 3^0$ surrounding
the epicenter of Tohoku earthquake (EQ), i.e., in the area
$N_{36.5}^{39.5} E_{141}^{144}$ by using the Japan Meteorological
Agency (JMA) earthquake catalogue. (Hereafter, the corresponding
magnitudes are labelled $M_{JMA}$.) Seismic moment $M_0$  and thus
the seismic energy was obtained from the moment magnitude $M_w$ by
applying the following approximate formulae obtained by
\citet{TAN04}. Then $M_0$ was deduced through the relation  $M_0
\sim 10^{1.5M_w}$ \citep{KAN78}.

From these $\kappa_1$ values, Fig. 1 has been constructed showing
the probability distribution $P(\kappa_1)$ of $\kappa_1$ on
several dates just before the main shock. The calculation was made
for various $M_{JMA}$ thresholds. {These thresholds
have been selected, as it should for the study of seismicity
changes \citep[e.g.,][]{Huang06}, to exceed the value
$M_{JMA}=3.4$ above which \citet{PNAS13} found that the JMA
earthquake catalogue is complete by analyzing the data since 1
January 1984 until the $M_w$9.0 Tohoku earthquake occurrence
within the area $N_{25}^{46} E_{125}^{148}$ practically covering
the whole Japanese region.} Examples are given in Figs. 1(a), (b)
and (c), for $M_{JMA}
>$ 3.4, 3.9 and 4.8,  {where 113, 73 and 30 EQs have been used into the calculation}, respectively. The cyan
crosses depict $P(\kappa_1)$ at the end of February, i.e., at
00:38 LT on 27 February, while the magenta ``x'' shows
$P(\kappa_1)$ almost two hours after the $M$7.3 foreshock that
occurred at 11:45 LT on 9 March 2011. We observe that
$P(\kappa_1)$ was displaced in the latter to lower $\kappa_1$
values close to zero.  At later times, i.e., almost 9 (blue), 16
(green) and 21 (red) hours after the foreshock, the height of
$P(\kappa_1)$ close to zero became smaller and $P(\kappa_1)$
finally exhibited a local maximum at 0.07 around 08:40 LT on 10
March 2011 (see the black vertical arrows in the insets). The main
shock could have happened at any moment after this, as mentioned
in Section \ref{sec1}. It is remarkable that the condition
$\kappa_1=$0.07 was not observed, in a wide range of magnitude
thresholds, before the aforementioned $M$7.3 foreshock indicating
the system has not reached yet the critical point at that stage.

\section{Conclusions}\label{sec4}
By analyzing the seismicity of Japan in natural time,
\citet{PNAS13} have found that the fluctuations of the order
parameter $\kappa_1$ exhibited unprecedented minimum around 5
January 2011, i.e., almost 2 months before 11 March $M_w$9 Tohoku
EQ. This date is likely to coincide with the initiation of a SES
activity when considering the conclusions drawn \citep{TECTO13}
from earlier cases in Japan where both seismic data and
geoelectrical data were available. Actually, \citet{Xu2013} found
that the anomalous magnetic field variations started on 4 January
2011, i.e., approximately on the same date. Here, it is also found
that the $\kappa_1$ values of seismicity computed after this date
converged to $\kappa_1= 0.07$ at 08:40 LT on 10 March 2011, i.e.,
the day before the Tohoku EQ occurrence, thus signalling the
impending risk. Quite interestingly, these $\kappa_1$ values did
not converge to 0.07 before the 9 March $M$7.3 EQ, indicating that
the critical point was not reached at the stage of this foreshock.
This fact may provide the means of identifying whether an EQ is a
foreshock or a main shock when natural time analysis is employed.

In a separate publication (Sarlis, Skordas and Varotsos to be published), we draw attention to the following point motivated
by the aspects of the PSPC model: Changes in the correlation properties of other associated
physical quantities \citep{VAR11,EPL12} like crustal deformation
orientation \citep{VAR11} (by analyzing GPS measurements) have
been observed approximately on the same date(s).

%\section*{{Acknowledgements}}
{}

\begin{figure}\label{fig1}
\includegraphics[scale=0.7]{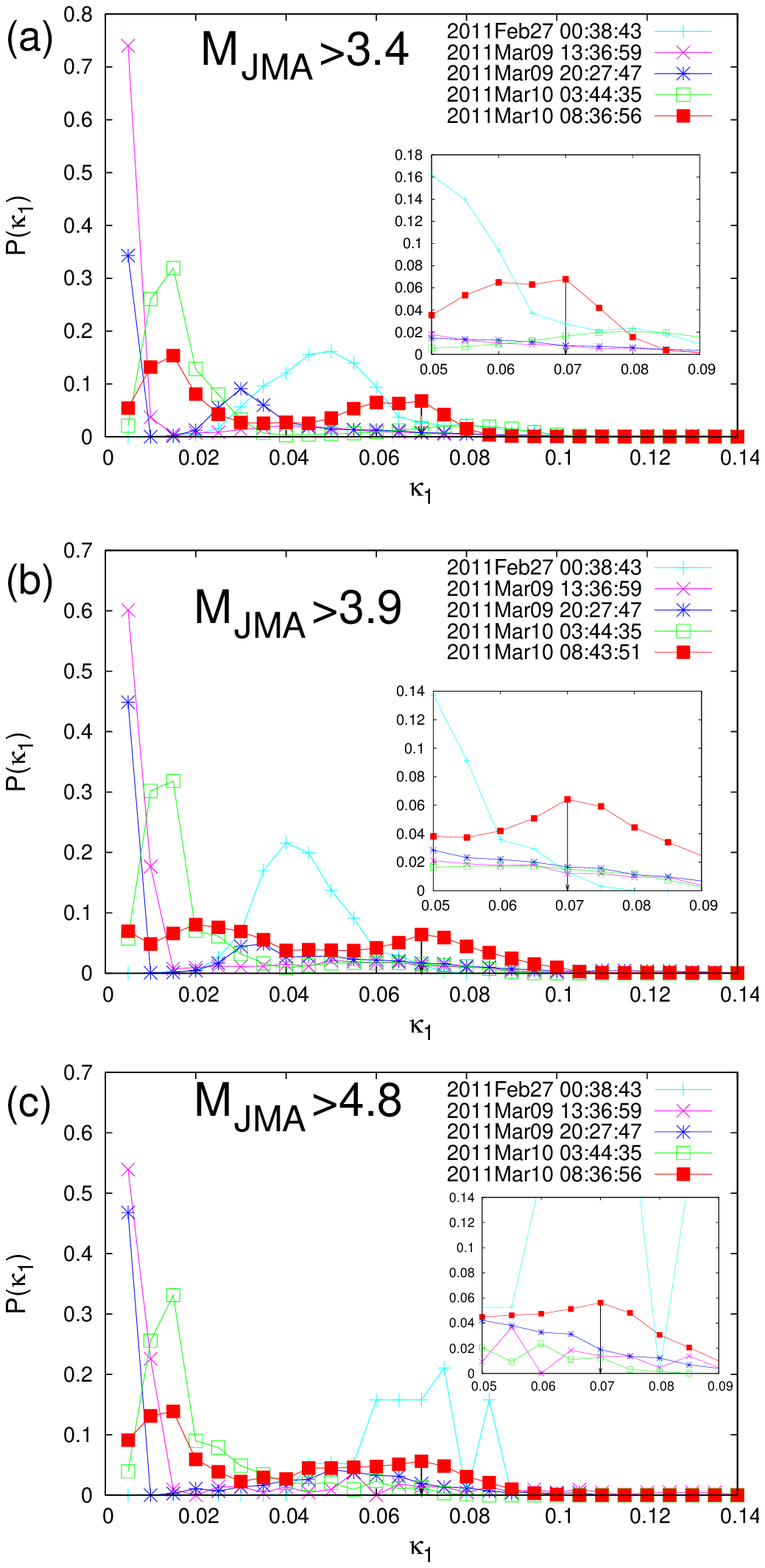}
\caption{(color) The probability $P(\kappa_1)$ of
finding a value in the range  $\kappa_1 \pm 0.0025$ as a function
of $\kappa_1$ for the seismicity after 5 January 2011. The values
of $\kappa_1$ have been found by studying all the possible
subareas within the area $N_{36.5}^{39.5} E_{141}^{144}$ for three
$M_{JMA}$ thresholds. The insets are enlarged excerpts around
$\kappa_1=$0.07. The dates and times are for the last EQ
considered in each calculation.}
\end{figure}

\end{document}